\documentclass[letterpaper]{article} 
\usepackage{aaai24}  
\usepackage{times}  
\usepackage{helvet}  
\usepackage{courier}  
\usepackage[hyphens]{url}  
\usepackage{graphicx} 
\usepackage{natbib}  
\usepackage{caption} 
\usepackage{algorithm}
\usepackage{algorithmic}

\usepackage{newfloat}
\usepackage{listings}
\DeclareCaptionStyle{ruled}{labelfont=normalfont,labelsep=colon,strut=off} 
\floatstyle{ruled}
\newfloat{listing}{tb}{lst}{}
\floatname{listing}{Listing}
\usepackage{xcolor}

\usepackage{xcolor}
\newcommand{\answerYes}[1]{\textcolor{blue}{#1}} 
\newcommand{\answerNo}[1]{\textcolor{teal}{#1}} 
\newcommand{\answerNA}[1]{\textcolor{gray}{#1}}

\usepackage{amsmath}
\usepackage{cleveref}
\usepackage{multirow}

\usepackage{enumerate}
\usepackage{enumitem}
\newenvironment{packed_item}{
\begin{itemize}[leftmargin=.1in]
  \setlength{\itemsep}{1pt}
  \setlength{\parskip}{0pt}
  \setlength{\parsep}{0pt}
}{\end{itemize}}
\title{Characterizing Online Criticism of Partisan News Media  \\Using Weakly Supervised Learning}
\author {
    Karthik Shivaram\textsuperscript{\rm 1},
    Mustafa Bilgic\textsuperscript{\rm 2}, 
    Matthew Shapiro\textsuperscript{\rm 3}, 
    Aron Culotta\textsuperscript{\rm 1}
}
\affiliations {
    \textsuperscript{\rm 1} Department of Computer Science, Tulane University, New Orleans, LA  ~ USA\\
    \textsuperscript{\rm 2} Department of Computer Science, Illinois Institute of Technology, Chicago, IL  ~ USA\\
    \textsuperscript{\rm 3} Department of Social Sciences, Illinois Institute of Technology, Chicago, IL  ~ USA\\
    kshvaram@tulane.edu, mbilgic@iit.edu, shapiro@iit.edu, aculotta@tulane.edu
}

\begin{document}

\maketitle

\begin{abstract}
We propose novel methods to identify tweets that criticize partisan news sources.  Prior work suggests that criticism, ridicule, and distrust  of news media all play important roles in hyperpartisanship, misinformation, and filter bubble formation. Thus, understanding the prevalence and temporal dynamics of media-targeted criticism can provide us with updated tools to assess the health of the information ecosystem. There is a scarcity of labeled data for this task, and we develop a weakly supervised learning approach that leverages multiple noisy labeling functions based on both the content of the tweet as well as the historical news sharing behavior of the user. Using this classifier, we explore how tweets expressing criticism vary by user, news source, and time, finding substantial spikes in media criticism during politically polarizing events, such as the investigation into Russian interference in the 2016 U.S.~elections and the 2017 ``unite the right'' rally in Charlottesville. This type of media-targeting criticism is also more likely to occur after users have been exposed to unreliable and hyperpartisan media. 

\end{abstract}

\section{Introduction}

The media is crucial for a functioning democracy~\cite{street2010, gentzkow2006}, but trust in news media has eroded. A recent Pew survey found that only 61\% of U.S.~adults have some or a lot of trust in the information they get from national news organizations, a drop of 15\% from just six years prior~\cite{pew22trust}. Such trends have implications for our political institutions, as distrust and animosity toward media contribute to hyperpartisanship, polarization, and misinformation sharing~\cite{rathje2021out,osmundsen2021partisan}. 
 
We currently lack methods to study how perceptions of the media evolve, how they vary according to news source, and how they are influenced by current events. Being able to observe media criticism --- historically and in real-time --- would allow us to monitor the health of the information ecosystem and identify trends in attitudes towards the media.

To address this methodological gap, this paper offers data and techniques to identify social media messages that express criticism and/or distrust of a news source (e.g., Figure~\ref{fig.tweet_distrust}). We collect over 3.5M tweets mentioning news sources over the past ten years and then train a neural network to categorize tweets based on their content as well as the user's past engagement behavior. Labeled data is scarce and costly to generate, so we apply a weak supervision approach, using noisy labeling functions based on keywords and user attributes to train the classifier. After validating the classifier on a smaller number of manually-labeled examples, we apply it to all of our historical data, analyzing the prevalence of media-critical tweets by user, news source, and over time. Given these advances in our understanding of news engagement dynamics, the primary contributions of this work can be summarized as follows:
\begin{figure}[t]
    \centering
    \includegraphics[width=.76\linewidth]{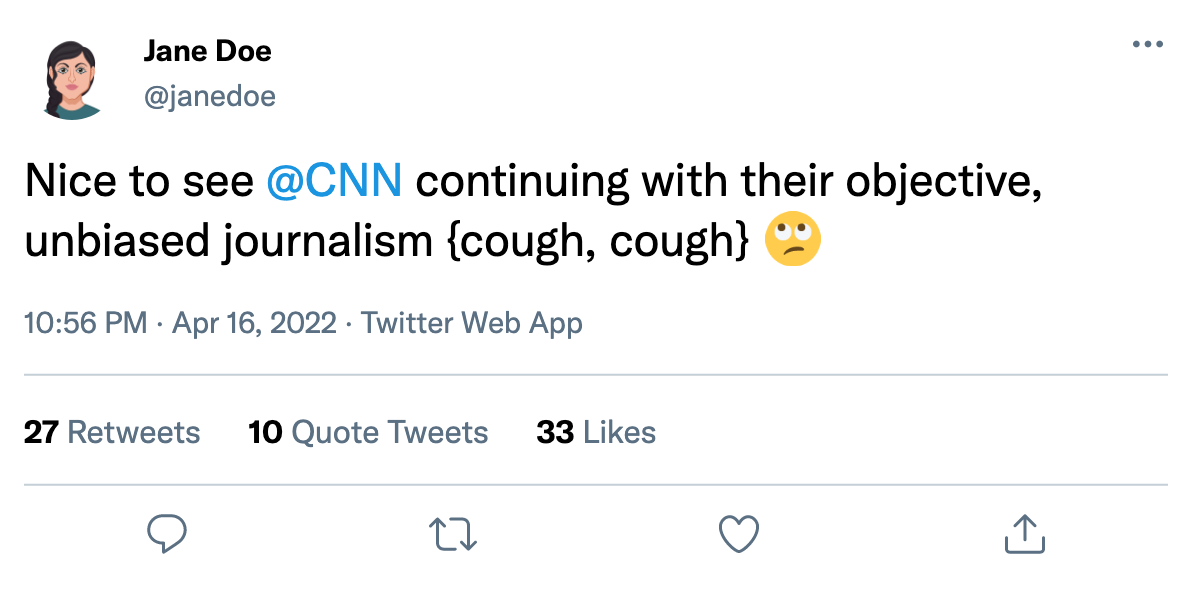}
    \caption{Example tweet critical of a news source.}
    \label{fig.tweet_distrust}
\end{figure}
\vspace{-.2cm}
\begin{packed_item}
\item \textbf{Dataset}: We construct a new dataset of 3.5M tweets that engage with one of 522 news sources over a ten year period. Tweet IDs, news sources, and inferred sharing intent are available at the paper repository.\footnote{\url{https://github.com/tapilab/icwsm-2024-news-criticism}}
\item \textbf{Weak supervision:} We find that classification using only text and user-based heuristics can provide accurate labels (89\% F$_1$) but with modest coverage (48\%). Training a weakly-trained classifier enables 100\% coverage while maintaining high accuracy (84\% F$_1$).
\item \textbf{Effects on Polarization Estimates:} We find that adjusting for media-criticising tweets provides a different picture of the diversity of user news engagement. Our accounting for news-critical tweets shows that users previously engaging with diverse news sources are actually more hyperpartisan.
\item \textbf{Criticism by user, news source, and over time:} After applying the classifier to the larger dataset, we find that the most criticised news sources are CNN and MSNBC on the left and Fox News and OANN on the right; that hyperpartisan users are more likely to post critical tweets; and that the rate of news criticism exhibited several significant spikes during key political events (e.g., during the investigation of Russian involvement in the 2016 U.S.~election).
\end{packed_item}

\section{Related Work}
\label{sec:related}

\paragraph{Political Stance Detection} Detecting media criticism may appear related to political stance and bias detection~\cite{demszky2019analyzing,darwish2020unsupervised,he2021detecting,li2021mean,dutta2022semi},
but it is a novel exercise. Media criticism goes beyond simply identifying the general political lean of a news article or topic, instead representing instances where messages are meant to ridicule, express distrust or sarcasm, or convey animosity \textit{towards a news source}. Thus, we must distinguish between a tweet criticising a politician who appears on CNN and a tweet that criticises CNN directly. To our knowledge, there has been no attempt to generate a classifier for detecting media criticism, which is essential given the implications of negatively framed content in social media~\cite{horner2021, ferrara2015, stieglitz2013}.

\vspace{-.1cm}
\paragraph{Media-Targeted Criticism} Journalism is scrutinized by the public \cite{carlson2016}, at times in a highly critical and formalized way.
The public expresses negative sentiment toward journalists and the institution of journalism itself with disgust and shame~\cite{shin2021}. This is particularly true for individuals who are least trustful of the news~\cite{karlsson2019}. Criticism of journalism is also motivated by efforts of extremist political groups that foster the view that the media is neither legitimate nor accurate~\cite{figenschou2019}. Identifying media-targeted criticism is essential when attempting to analyze factors that influence the spread of online misinformation \cite{gabriel2022misinfo}. It also helps us better understand the political lean of online media \cite{stefanov2020predicting}, functioning as a more nuanced tool for assessing media-based ideology.

\vspace{-.1cm}
\paragraph{Weakly Supervised Classification}
Weakly supervised learning, a sub-field of machine learning, uses noisy sources of supervision to minimize human annotation efforts. Prior work applies approaches such as self-training \cite{karamanolakis2021self}, co-training \cite{karamanolakis-etal-2019-leveraging}, and crowd-sourcing \cite{li2019exploiting}. 
Other work uses multi-task learning~\cite{ratner2018snorkel,ratner2019training} or generative models~\cite{yu2022learning,bach2017learning}  to reduce the effects of label noise.

\begin{table}
\small
\centering
  \begin{tabular}{rrr}
  \hline
\textbf{Type}                    & \textbf{Count} & \textbf{News Engagement}\\
\hline
All Tweets                     & 36,543,574  & 3,491,270 ~~(9.5\%)~~   \\
Quotes                     & 1,417,012   & 171,403 (12.1\%)~~   \\
Retweets                   & 18,823,632  & 1,965,171 (10.4\%)~~    \\
Replies                    & 8,495,253   & 670,138 ~~(7.9\%)~~    \\
Status                     & 7,807,677   & 684,558 ~~(8.8\%)~~    \\
\hline
\end{tabular}
  \caption{Tweets collected from 5,470 users and the fraction that reference one of 522 news sources.}
  \label{tab:data-col}
\end{table}

\section{Data}
\label{sec:data}
Our goal was to collect news engagement tweets that are (a) diverse with respect to the partisan lean of the news source, (b) diverse with respect to the partisan lean of the users, and (c) posted over many years in order to observe long-term trends both at the user level and in aggregate. To do so, we identified a diverse set of Twitter users who engage with political news using the following steps:

\vspace{-.2cm}
\paragraph{Step 1: Collect news sources.} We collected 419 English-based news sources from \textit{allsides.com}, a media rating site used often in prior work~\cite{baly-etal-2020-detect,liu2021interaction,sinno-etal-2022-political,lyuetal2022}. Each news source is associated with a partisan stance in $\{-2, -1, 0, +1, +2\}$, ranging from extreme liberal ($-2$) to extreme conservative ($+2$). To cover a more diverse range of media quality, we extend the above collection by including 103 low-reliability sources from \citet{osmundsen2021partisan}, itself originally sourced from \citet{guess2019fake} (42 sources) and \citet{grinberg2019fake} (125 sources).\footnote{We took 167 from \citet{osmundsen2021partisan} and dropped those that either (a) were already in AllSides or (b) did not have a Twitter account or website. This reduced the list to 103 sources.} These are rated as being either \textit{pro-Republican} or \textit{pro-Democrat}. We denote the partisan lean of these sources as -3 (pro-Democrat) and +3 (pro-Republican). We represent fake news with +3 and -3 stances to capture the extreme polarization attributed to such news, as it is considered a more intensified form of partisan bias, appealing strongly to users with corresponding political leanings. This representation aligns with polarization theory, which suggests that fake news is more attractive to polarized users due to its extremity and alignment with their partisan preferences (\citet{osmundsen2021partisan}).
We retrieved the Twitter handle and URL for each of the 522 resulting news sources.

\vspace{-.2cm}
\paragraph{Step 2: Identify users.} To identify users who engage with these 522 news sources, we used the Twitter Search API to find tweets that either mention a news source's Twitter account or contain a URL matching the news source's web domain. We submitted queries for each news source in Fall 2021, yielding 1.67M matching tweets.

While these matched users are likely to be actively engaged with news, we also wanted to widen the user set to account for a more diverse set of users. To do so, we used the Twitter Streaming API to sample from all English-language tweets posted during the same time period. We added these 59K tweets to the 1.67M already collected.

\vspace{-.2cm}
\paragraph{Step 3: Filter users.} For studying long term trends, we retained only those users identified in the previous step with accounts at least five years old. Additionally, to limit the impact of automated accounts, we removed users that appear to exhibit automated behavior based on frequency of tweets, number of followers, and number of friends (c.f., Appendix~\ref{sec:appendix_bot}). After filtering, we then sampled users to diversify by partisan stance and news source, ensuring that the dataset is not dominated by a handful of news sources. To do so, we sampled $\sim$600 users from each partisan stance (based on the news source they mention); within each partisan stance, we sampled an equal number of users for each news source. We added to this set a random sample of 1,200 users identified from the Streaming API in the previous step. This resulted in a final set of 5,470 users representing a diverse set of political interests and engagement.

\vspace{-.2cm}
\paragraph{Step 4: Collect and annotate timelines.} After the users were sampled, we collected each user's entire timeline to identify a larger set of news engagement tweets. From the 5,470 users, we collected nearly 37M tweets spanning ten years. For each tweet, we searched for a mention or URL that refers to any of the 522 news sources identified in Step 1 and labeled each matching tweet with its corresponding partisan score (i.e., the score of the referenced news source). Of the 37M tweets, we observe in Table~\ref{tab:data-col} that 3.5M engage with at least one of the 522 news sources, suggesting that these users are, by design, quite engaged with political news and thus should not be considered representative of all Twitter users or the U.S. population as a whole. (Please see \S\ref{sec:limitations} for more discussion of such limitations.)

\section{Problem Formulation}\label{sec:problem}
With the above data, we express our problem as follows: For each tweet that mentions a news source, we must determine whether or not that tweet is critical of the news source. A \textit{critical} tweet is one that expresses disapproval of the media source. Note that this is distinct from tweets that criticize entities described in a news article, as well as from tweets that criticize the event the article describes. We use the term \textit{critical} to encompass a variety of connotations, such as ridicule, distrust, animosity, and sarcasm. As usual, these expressions range from the direct (``@FoxNews is garbage.'') to the subtle (``Nice to see @CNN continuing with their objective, unbiased, journalism.  \{cough, cough\}'').

We formulate this as a binary classification task. For each tweet $t_i$ mentioning a news source, we assign a class label $y_i \in \{0,1\}$, where $y_i=1$ indicates that the author is criticising the news source, and $y_i=0$ indicates the absence of criticism.

Based on our initial exploration of the data, we make the following simplifying assumptions to formulate a more tractable task: (1) we remove direct retweets of news sources, as these do not add any additional context to assess intent (e.g., ``RT @CNN: breaking news ...'') ; (2) we remove tweets that are part of threaded replies, as it is challenging to determine who the target of criticism is (e.g., ``@JoeSmith @CNN You are garbage''); and (3) we restrict analysis to tweets that either reply directly to a news source (e.g., ``@FoxNews \#FakeNews'') or mention the news source in the body of the tweet (e.g., ``When will @CNN stop lying?'').

Finally, in line with prior work showing that engagement occurs most frequently with ideologically extreme content \cite{eady2021}, we discovered early on that the rate of criticism is significantly higher for more partisan news sources. To focus on the most salient subset of data, we thus target the 216 news sources with partisan stance in $\{-3, -2, +2, +3\}$. These sources include popular outlets such as CNN, MSNBC, and Slate on the left, and Fox, OANN, and Breitbart on the right. With these assumptions, our final sample consists of 1.2M tweets that qualify for classification.

\section{Methods}
\label{sec:methods}
Given the lack of labeled data, and given the presence of several strong classification signals based on user and keyword features, weakly supervised learning provides an efficient methodology to train a classification model for this task~\cite{kim2012bayesian,ratner2018snorkel,li2019exploiting}. The overall approach is to first define a set of \textit{labeling functions} that can provide noisy labels for a large subset of data. For example, the presence of the term \#FakeNews may serve as a strong labeling function. As well, when a user who mostly engages with strongly conservative media mentions a strongly liberal news source, it is probable that the intent is to criticise the liberal news source.

We train a classifier instead of using these labeling functions directly because a classifier can generalize beyond the specific rules encoded in the labeling functions, allowing it to make predictions on unseen data with greater performance, also a classifier can integrate the signals from multiple labeling functions, potentially resolving conflicts and inconsistencies among them to provide more robust and reliable predictions.

Once these labeling functions are defined, they are used to create training data for a classifier. To account for label noise, we compare several weakly supervised learning methods designed for such scenarios. We next describe the labeling functions and classification methods in turn.

\subsection{Labeling Functions}

We provide here an overview of the labeling functions used to train the classifier. (For more details, please see Appendix \ref{sec:appendix_lf}.) We define a labeling function $\phi$ as a collection of heuristics that maps a given tweet to a corresponding label $y \in \{0,1,-1\}$. 
For our purposes, $0$ represents the absence of criticism, $1$ represents the presence of criticism, and $-1$ represents the inability of the labeling function to assign labels (abstention) due to either missing information or certain cut-off thresholds not being met, as described below. We implement labeling functions based on the three following types of information:

\vspace{-.2cm}
\paragraph{User features ($\phi_{up}$):} This labeling function relies on a user's historic news engagements and the political accounts they follow. First, we estimate the partisan stance of the user (conservative or liberal) based on whom they follow and the partisan lean of the news sources with which they engage. For example, if more than 90\% of the political accounts they follow are liberal, and if more than 90\% of their news engagement tweets are liberal, the user is labeled as liberal.\footnote{The 90\% threshold is a tuning parameter to tradeoff precision and coverage.} This labeling function annotates each news engagement tweet that is aligned with the user's partisan lean as 0. Tweets of the opposite partisan lean are annotated as 1. This function abstains for tweets from users for whom we could not infer partisan lean based on the thresholds above. 

\vspace{-.2cm}
\paragraph{Text Features ($\phi_{tt}$):} This labeling function relies on the text of the tweet in which the user mentions a news source. We consider keywords both indicative of criticism (e.g., ``propaganda,'' ``fake news'') as well as those indicative of support (e.g., ``must watch'', ``worth reading''). In addition to individual words/phrases, we also consider word collocations --- e.g., when ``false'' and ``story'' appear in any order, the text is labeled as critical.

\vspace{-.2cm}
\paragraph{Union of the above ($\phi_{un}$):} This labeling function takes the union of the prior two functions $\phi_{up}$ and $\phi_{tt}$, ignoring conflicting assignments. That is, if the two functions agree, or if one of them abstains, the predicted label is returned; otherwise, it abstains. By removing conflicting labels, we expand the coverage of the single labeling functions and reduce label noise.

To assess the coverage of each function, we apply them to the unlabeled data filtered as described in \S\ref{sec:problem} and report the estimated label distribution in Table~\ref{tab:lf_stats}. We observe that $\phi_{tt}$ and $\phi_{up}$ exhibit similar levels of coverage, labeling 30\% and 26\% of the data, respectively, and each labeling 3\% of the data as critical. As a fraction of the labeled instances, excluding abstentions, each method labels about 11\% as critical. The union function $\phi_{un}$ improves coverage to 48\%, while also assigning about 11\% of the labeled instances to the critical class. This limited overlap between the instances labeled by the user and text labeling functions suggests that they are complementary. We discuss the accuracy of these labeling functions on manually annotated data in \S\ref{sec:experiments}.

\subsection{Classification Models}

In this section, we discuss the different models trained based on the labeling functions from the previous section. We consider separate neural networks based only on user features or only on text features, as well as a network that combines the two. Unlike the labeling functions, these models are binary classifiers: critical vs.~not critical.

\vspace{-.2cm}
\paragraph{User Network}
This network model takes as its input hand-crafted features based on the user's Twitter profile as well as their historic news engagements. These include features such as the distribution of partisan stances among a user's mentions or follows, the partisan stance of the tweet being classified, and how the news source is referenced (e.g., direct reply or mention in tweet body).\footnote{A complete list of features is available in Appendix ~\ref{sec:user_features}.}

We use a simple fully-connected network for this model. For each tweet, we extract the above features, $f_i$, and pass them through one hidden layer with relu activation, followed by a classification layer with sigmoid activation:
 \begin{equation*}
     z_u = \hbox{relu}(W_uf_i+b_u) ~~~~~~ \hat{y_i} = \sigma(W_oz_u+b_o)
 \end{equation*}

\vspace{-.2cm}
 \paragraph{Text Network}
This network uses the actual text of the tweet (and any referenced tweet) to perform classification. To improve generalizability, we first pre-process all tweets by replacing Twitter handles with a placeholder token. We then pass each tweet through a version of the RoBERTa language model pre-trained on English tweets \cite{barbieri2020tweeteval} to obtain word-level representations $\{a_0,a_1,...a_p\}$, after which we perform a pooling aggregation to obtain a single vector representation $r_i$. This is passed through one hidden layer and one classification layer with sigmoid activations:
 \begin{equation*}
     z_t = \sigma(W_tr_i+b_t) ~~~~~~~  \hat{y_i} = \sigma(W_gz_t+b_g)
 \end{equation*}

\vspace{-.2cm}
\paragraph{Combined Network}
This network combines user features and text based representations together in order to identify if a given tweet contains the presence of criticism towards an engaged news source. For each tweet we obtain the intermediate representations $z_u$ (from the User Network) and $z_t$ (from the Text Network) and pass them through linear layers to obtain $z_{cu}$ and $z_{ct}$. These are then concatenated to obtain $z_c = [z_{cu},z_{ct}]$ and passed through to the final output layer to compute the corresponding class label $\hat{y_i}$: 
 \begin{equation*}
     z_{cu} = \sigma(W_{cu}z_u+b_{cu}) ~~~~     z_{ct} = \sigma(W_{ct}z_t+b_{ct})
 \end{equation*}
 \begin{equation*}
     \hat{y_i} = \sigma(W_hz_c+b_h)
 \end{equation*}
We use binary cross-entropy as the objective function to train all networks.

\subsection{Label Denoising}

Given the label noise inherent in the labeling functions above, we additionally experiment with several label denoising approaches. The general pipeline consists of fitting a probabilistic  model that combines the labels generated by different labeling functions and denoises them to return soft (weighted) labels~\cite{zhang2021wrench}. These soft labels are then used to train our classification models. We consider four different approaches that are appropriate for our task: Dawid Skene (DS)~\cite{dawid1979maximum}, IBCC~\cite{kim2012bayesian}, EBCC~\cite{li2019exploiting}, and Data Programming (DP)~\cite{ratner2018snorkel}.\footnote{While other methods exist, they require a larger number of labeling functions to be effective.} These methods leverage agreements and conflicts between the different labeling functions in order to reduce label noise. For example, if two heuristics predict that a tweet is critical but a third does not, these denoising methods would learn how to resolve conflicts based on the frequency of disagreements for each heuristic. We use both $\phi_{up}$ and $\phi_{tt}$ generated labels to train these models, and we use the implementations provided by WRENCH~\cite{zhang2021wrench} (with default hyperparameters), a weak supervision benchmark library for classification tasks.

\section{Classification Experiments}\label{sec:experiments}
In this section we describe the experiments to validate the classification approach. To construct a smaller dataset for tuning and validation, we first randomly sample 300 tweets from the full dataset and manually annotate them. Because there is high class imbalance (only about 10\% critical), this resulted in only a small number of critical examples. Thus, we augment these data by using our labeling functions to identify a sample of critical and non-critical examples, which we then manually annotate. After this annotation process we sub-sample from this resulting set to have balanced label distributions, splitting them into validation and test datasets, ensuring that there are no overlapping users across our training, testing, and validation sets. The final dataset sizes for test and validation are, respectively, 312 and 233.  The inter-annotator agreement of two annotators (one an author and another an unaffiliated student) has a Cohen's kappa of 0.71, indicating substantial agreement.

\begin{table}[t]
\centering
\small
\begin{tabular}{cccc}
\hline
\textbf{function} & \textbf{critical} & \textbf{not critical} & \textbf{abstain}\\
\hline
$\phi_{tt}$   & 5,087 (3\%)    & 39,356 (27\%)     & 103,594 (70\%)\\
$~\phi_{up}$   & 4,600 (3\%)    & 34,601 (23\%)     & 108,836 (74\%)\\
$~\phi_{un}$   & 7,872 (5\%)    & 63,181 (43\%)     & ~~76,984 (52\%)\\ 
\hline
\end{tabular}
\caption{Labeling function output on unlabeled data}
  \label{tab:lf_stats}
\end{table}

\begin{table}[t]
\small
\centering
\begin{tabular}{cccccc}
\hline
\textbf{Function} & \textbf{Coverage} & \textbf{F1} & \textbf{Prec} & \textbf{Rec} & \textbf{Acc} \\ 
\hline
$~\phi_{up}$ & 0.388               & 0.845      & 0.862             & 0.853          & 0.860            \\
$\phi_{tt}$ & 0.391               & 0.869      & 0.879             & 0.869          & 0.869            \\
$~\phi_{un}$ & 0.600               & 0.888      & 0.888             & 0.888          & 0.888            \\ \hline
\end{tabular}
\caption{Labeling function accuracy on test data.}
  \label{tab:lf_test}
\end{table}

\begin{figure}[t]
    \centering
    \includegraphics[width=0.99\linewidth]{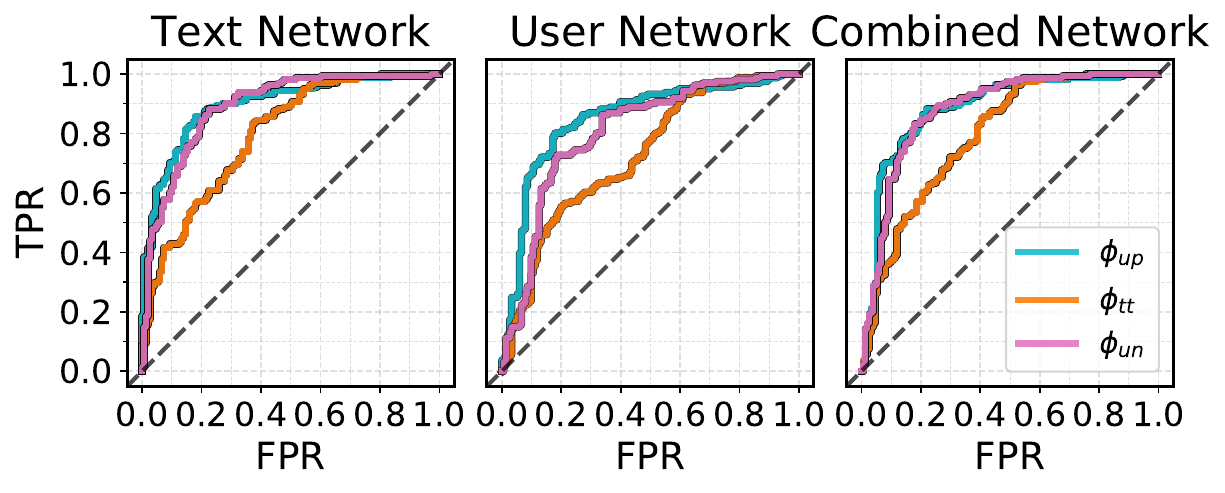}
    \caption{ROC curves of network models trained using different labeling functions.}
    \label{fig.roc_comp}
\end{figure}

\begin{table}[t]
  \small
  \centering
  \begin{tabular}{lc}
  \hline
  \multicolumn{1}{c}{\textbf{Network}}  & \textbf{ROC AUC} \\ \hline
  Text Network + $\phi_{up}$            & 0.800 $\pm$ 0.028 \\
  Text Network + $\phi_{tt}$            & 0.719 $\pm$ 0.008 \\
  Text Network + $\phi_{un}$            & 0.810 $\pm$ 0.017 \\
  Text Network + DS                     & \multicolumn{1}{l}{\textbf{0.840 $\pm$ 0.007}} \\
  Text Network + IBCC                   & \multicolumn{1}{l}{0.822 $\pm$ 0.014} \\
  Text Network + EBCC                   & \multicolumn{1}{l}{0.836 $\pm$ 0.006} \\
  Text Network + DP                     & \multicolumn{1}{l}{0.812 $\pm$ 0.013} \\
  User Network + $\phi_{up}$            & 0.749 $\pm$ 0.049 \\
  User Network + $\phi_{tt}$            & 0.657 $\pm$ 0.022 \\
  User Network + $\phi_{un}$            & 0.744 $\pm$ 0.025 \\
  Combined Network + $\phi_{up}$        & 0.810 $\pm$ 0.015 \\
  Combined Network + $\phi_{tt}$        & 0.723 $\pm$ 0.006 \\
  Combined Network + $\phi_{un}$        & 0.796 $\pm$ 0.008 \\
  Combined Network + DS                 & \multicolumn{1}{l}{0.816 $\pm$ 0.015} \\
  Combined Network + IBCC               & \multicolumn{1}{l}{0.784 $\pm$ 0.038} \\
  Combined Network + EBCC               & \multicolumn{1}{l}{0.810 $\pm$ 0.023} \\
  Combined Network + DP                 & \multicolumn{1}{l}{0.826 $\pm$ 0.015} \\ \hline
\end{tabular}
\caption{Test set ROC AUC for combinations of model, labeling function, and label denoising methods.}
\label{tab:model_perf}
\end{table}

To train each weakly supervised model, we apply our labeling functions ($\phi_{up}, \phi_{tt}, \phi_{un}$) to the unlabeled tweets filtered according to \S\ref{sec:problem}. Then, we sample a balanced distribution of labels across critical and non-critical classes for each of the three labeling functions. The final number of weakly supervised training examples used for each labeling function is 9,200 
for $\phi_{up}$, 10,174 for $\phi_{tt}$, and 17,962 for $\phi_{un}$. We train all possible combinations of our different network models and labeling functions, using the Adam \cite{kingma2014adam} optimizer. We tune each network across different hyperparameter values (see Appendix, Table~\ref{tab:hyper}) and select the best parameters based on validation accuracy to identify our best models. These networks are implemented in Pytorch \cite{paszke2019pytorch}.

Table~\ref{tab:lf_test} reports the performance of our labeling functions alone on the test set. We observe that the labeling functions are rather reliable (.845-.888 F$_1$ for the subset that they are able to classify), with the text heuristics slightly more accurate than the user heuristics; both have modest coverage ($\sim$40\% of instances are able to be annotated by each labeling functions). The union function offers an apparent improvement over both, increasing coverage by $\sim$20\% and accuracy by 1-2\%.\footnote{The coverage is higher here than in Table~\ref{tab:lf_stats} since the manually labeled data includes both uniformly sampled tweets as well as those annotated by the labeling functions.}

\begin{figure*}[t]
    \centering
    \includegraphics[width=.95\linewidth]{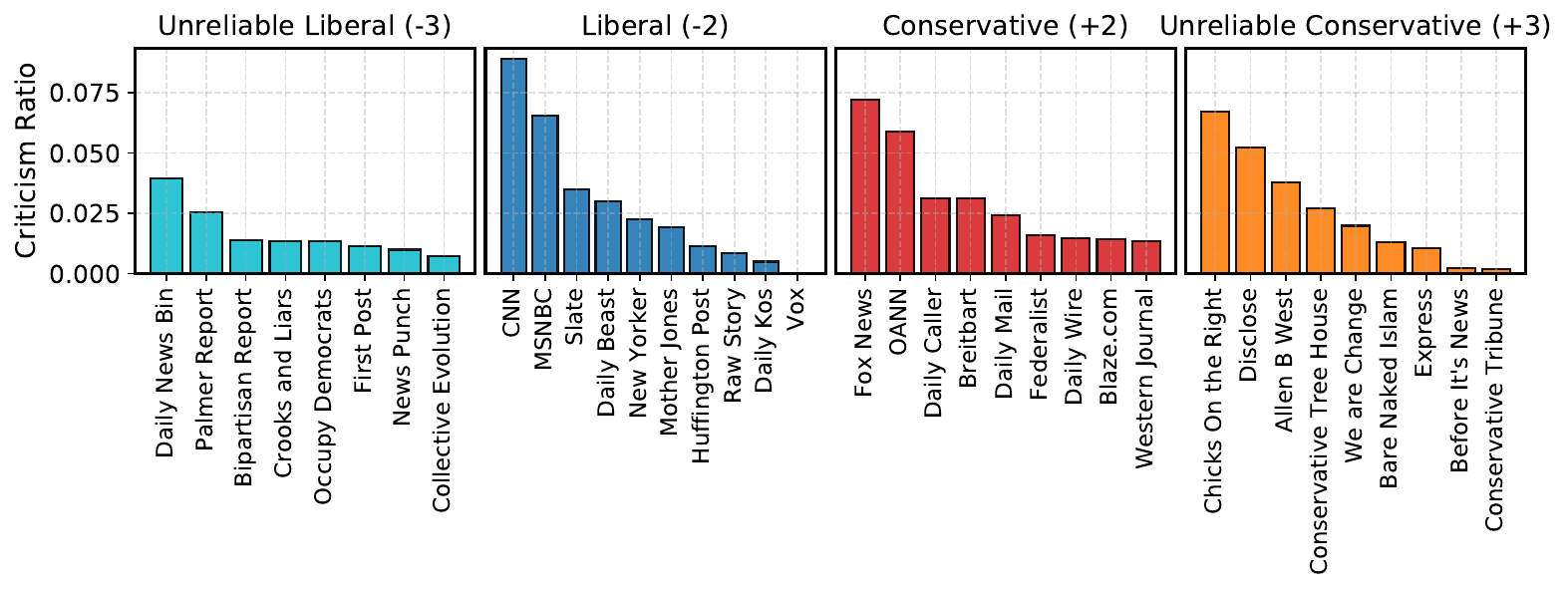}
    \caption{Criticism shown towards the most mentioned news sources from each partisan stance.}
    \label{fig.distrust_ns}
\end{figure*}
\begin{figure}[t]
    \centering
    \includegraphics[width=0.95\linewidth]{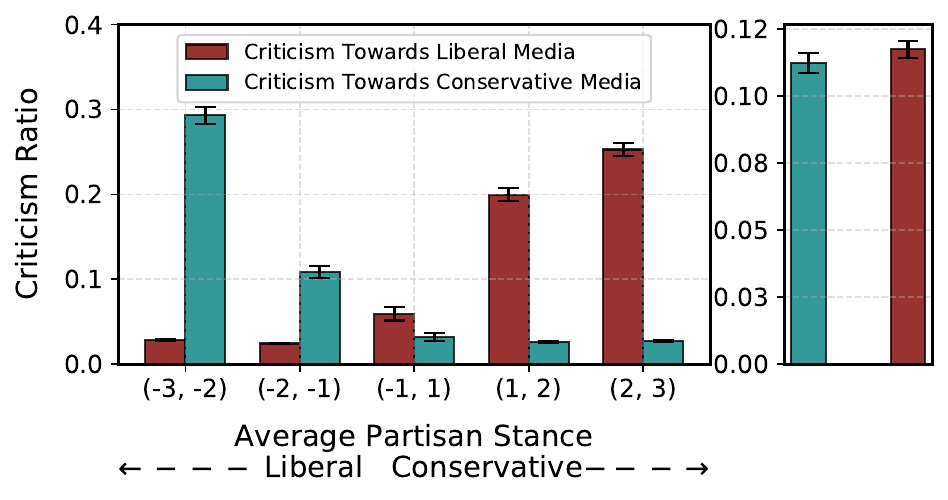}
    \caption{Criticism ratio by partisan stance of the user (left panel) and in aggregate (right panel).}
    \label{fig.dratio_user}
\end{figure}

Turning to the classification models, Figure \ref{fig.roc_comp} shows the ROC curves for each model/labeling function combination. 
We see that the text and combined networks perform better than the user network at different threshold values and across all three labeling functions. The worst performing labeling function across all three networks is $\phi_{tt}$, which may be due to the poor generalization of the keyword-based heuristics. 

Table \ref{tab:model_perf} shows each classification model's ROC AUC score along with the Label Denoising enhancements.
We observe that the text and combined networks are comparable in performance across different training settings. We also see that using soft-labels generated by the different label models (\textit{Dawid Skene, IBCC, EBCC,} and \textit{Data Programming}) improves performance compared to just training the models with hard labels generated by a single labeling function. The best performing model is the text network that uses soft-labels predicted by the \textit{Dawid Skene} label model, achieving an ROC AUC score of 0.840, averaged across five random seed settings. This is a 3\% improvement over using the text network on the $\phi_{un}$ labeling function alone.

\section{Analysis of Media-Targeted Criticism}
\label{sec:analysis}

As our main interest for this work is to characterize criticism shown towards partisan news media, we next use our best performing model to classify all data filtered according to \S\ref{sec:problem}. The resulting
dataset contains $\sim$1.2 million tweets, of which $\sim$1.16 million were labelled as non-criticism and $\sim$45K were labelled as criticism. (Appendix \ref{sec:validation} reports additional validation measures for these classifications.)

\subsection{Criticism by User Partisan Stance}
\label{sec:partisan_stance}
To analyze how criticism varies across users with different partisan preferences, we bin all users into five bins based on their average partisan stance, which is estimated by the stance of news sources a user engages with through direct retweets.\footnote{We assume direct retweets is a reliable indicator of user support for a particular news source.} For each bin, we compute the criticism ratio, defined as the proportion of news engagement tweets that criticise a news source. The results in Figure~\ref{fig.dratio_user} show that users with extreme partisan preferences (Bins 1 and 5) are much more likely to express criticism than users with more moderate preferences. Another interesting observation is that users that are moderately liberal (bin 2) exhibit less criticism compared to moderate conservatives (bin 4). The rightmost panel of Figure~\ref{fig.dratio_user} shows the overall criticism ratios, ignoring user bins, indicating a slightly higher level of criticism towards liberal media than towards conservative media, though the differences do not appear to be significant.

Note that Figure~\ref{fig.dratio_user} also indicates a small rate of ``self-criticism'' --- e.g., the leftmost bar indicates that a small ($<.04$) fraction of tweets from strongly liberal users about liberal media is critical; the same is true regarding conservative users and conservative news. While some of this may be due to unavoidable classification errors, a manual inspection confirmed that many of these are accurate. For example, many far right conservatives criticised Fox News for not supporting the discredited theory that the election was rigged: ``@FoxNews Please tell Chris Wallace there is serious fraud and the election is not even close to being decided! You folks are just hacks now, especially CW, not fair and balanced and unafraid that’s for sure.''

\subsection{Criticism by News Source}
Figure~\ref{fig.distrust_ns} plots the criticism ratio for the top ten most mentioned news sources from each partisan stance. Among liberal news media, CNN receives the highest rate of criticism ($\sim$9\% of all engagements), followed by MSNBC ($\sim$6\%). For conservative media, Fox ($\sim$7\%) is targeted the most, followed by OANN ($\sim$6\%). We also note that criticism shown towards unreliable conservative sources is greater than that shown towards unreliable liberal sources, although overall engagement with +3 news sources relative to -3 news sources is also greater.

The results suggest that more popular news sources receive higher rates of criticism, consistent with a recent Pew survey of media distrust~\cite{pew20media}, which estimates rates of cross-partisan distrust in each news source. The rankings of news sources in that survey align with the criticism ratios we find here: Fox (61\% of liberals distrust), Breitbart (36\%), DailyCaller (9\%); CNN (58\% of conservatives distrust), MSNBC (47\%), Huffington Post (34\%), Vox (11\%). Thus, even though some sources may be seen as more extreme (e.g., DailyCaller vs. FoxNews; MSNBC vs. CNN), their lower popularity results in lower rates of criticism.

\subsection{Effect of Criticism on Diversity Measures}

Identifying tweets containing media-targeted criticism may affect estimates of user polarization~\cite{conover2011political,garimella2017long,garimella2021political}. For example, a user who posts supportive tweets for liberal news sources and critical tweets for conservative news sources may appear to have diverse news consumption if the intent of the tweet is ignored. We thus examine how measures of user diversity change after removing critical tweets.

To measure the diversity of a user's news engagements, we employ the normalized stance entropy measure (\textbf{NSE}) used in prior work on filter bubbles~\cite{liu2021interaction}:
$\text{NSE} = \frac{-\sum_{i=1}^{m}{p_i{\log p_i}}}{\log m} $,
where $p_i$ is the fraction of a user's engagements that belong to a particular stance $i \in \{-3,-2,2,3\}$, and $m=4$ is the total number of partisan stances. NSE has a maximum value of 1, and higher values indicate more diverse news engagement.

\begin{figure}[t]
    \centering
    \includegraphics[width=0.7\linewidth]{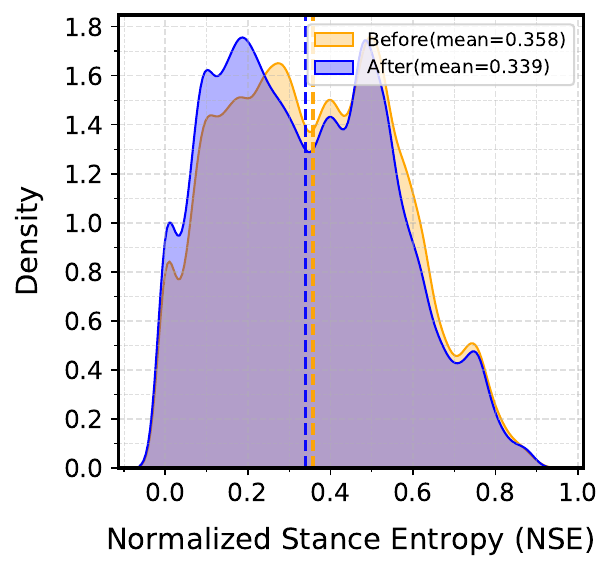}
    \caption{Comparison of Normalized Stance Entropy before and after removing critical tweets.}
    \label{fig.stance_entropy}
\end{figure}

Figure~\ref{fig.stance_entropy} shows the distribution of NSE scores before and after removing critical tweets. We can see that, after removing critical tweets, NSE drops from .358 to .339 on average (statistically significant at $p<.0001$ according to a $t$-test). The most noticeable change is the reduction in users in the middle range (.25 to .5), and a corresponding 13\% increase in the number of users with low diversity (0 to .25). These results indicate that measures of user diversity do decrease once critical tweets are accounted for.

\begin{figure}[ht]
    \centering
    \includegraphics[width=.99\linewidth]{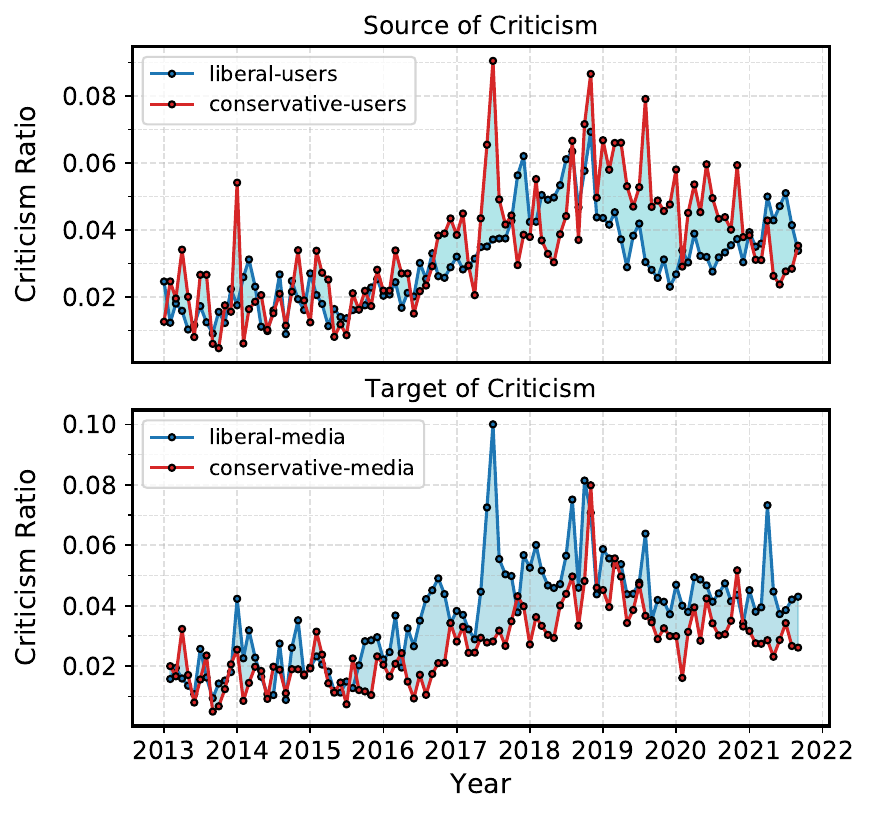}
    \caption{Criticism across time by partisan stance of the user (top panel) and news source (bottom panel).}
    \label{fig.distrust_time}
\end{figure}

\subsection{Criticism Over Time}

To investigate how criticism towards news media changes over time, we calculate the fraction of critical tweets per month. Figure~\ref{fig.distrust_time} plots both the source of criticism (top panel) and the target of criticism (bottom panel). We observe that criticism towards media has increased over time, with liberal media receiving higher criticism ratios than conservative media for most time periods. 

We also observe substantial spikes in these time series (mid-2017, late-2018 and mid-2019). To identify possible events corresponding to these spikes, we extract the most common terms used during these time windows. For the mid-2017 spike, prominent terms like \{\textit{Trump, President, Obama, Comey, Russia, police, Charlottesville, Syria, Muslim}\} reference events such as the investigations into Russian involvement in the 2016 U.S. elections, the ``unite the right'' rally in Charlottesville, and the Syrian War.  For the 2018 spike,  terms like \{\textit{Kavanaugh, vote, women, Mueller, investigation}\} refer to the sexual misconduct allegations against Supreme Court judge Brett Kavanaugh and  Robert Meuller's investigation into Russian interference in the 2016 U.S. election. For the mid-2019 spike, terms like \{\textit{Mueller, racist, 2020, Biden, children, border, Epstein}\} refer to the detention of child migrants during the border crises and the arrest of Jeffrey Epstein for sex trafficking crimes. We thus find that certain high-profile, hyper-partisan events can lead to spikes in media criticism.

\subsection{Progression Towards Criticism}
Finally, we study how individual users engage with news over time, specifically the context surrounding the first occurrence of a critical tweet. To assess whether this takes place before or after users engage with partisan/unreliable media, we separate users into liberal and conservative groups based on the stance of their direct retweets (as in \S\ref{sec:partisan_stance}).  For liberal users, we consider all who have engaged with -3 and -2 sources as well as those who have posted at least one tweet critical of conservative media (denoted $C_C$). Analogously, for conservative users, we consider those who have engaged with +3 and +2 sources while posting at least one tweet critical of liberal media (denoted $C_L$). We then find the first occurrence of each engagement type for each user and count the frequency of each sequence. The results in Table~\ref{tab:prog_seq} indicate that, for both sides, the most common sequence is partisan $\rightarrow$ unreliable $\rightarrow$ critical. The next most-common sequence is partisan $\rightarrow$ critical $\rightarrow$ unreliable. 
These results suggest that engagement with low reliable, hyperpartisan media often precedes media criticism, though a more rigorous study is required to isolate causal effects.

\begin{table}[t]
\small
\centering
\begin{tabular}{l @{\hspace{-1ex}} c @{\hspace{-1ex}} c}
\hline
\multicolumn{1}{c}{\textbf{User Group}} & \multicolumn{1}{c}{\textbf{Sequence}} & \multicolumn{1}{c}{\textbf{\% of Users}} \\ \hline
\textbf{Liberal}                        & $-2 \rightarrow -3 \rightarrow C_C$                           & \textbf{49.07}                           \\
                                        & $-2 \rightarrow C_C \rightarrow -3$                        & 45.91                                    \\
                                        & $C_C \rightarrow -2 \rightarrow -3$                          & 2.77                                     \\
                                        & $-3 \rightarrow -2 \rightarrow C_C$                          & 2.11                                     \\
                                        & $C_C \rightarrow -3 \rightarrow -2$                         & 0.13                                     \\
                                        & $-3 \rightarrow C_C \rightarrow -2$                          & 0.00                                       \\ \hline
\textbf{Conservative}                   & $2 \rightarrow 3 \rightarrow C_L$                            & \textbf{46.33}                           \\
                                        & $2 \rightarrow C_L \rightarrow 3$                            & 33.79                                    \\
                                        & $3 \rightarrow 2 \rightarrow C_L$                            & 13.03                                    \\
                                        & $C_L \rightarrow 2 \rightarrow 3$                           & 6.07                 \\
                                        & $3 \rightarrow C_L \rightarrow 2$                            & 0.49                     \\
                                        & $C_L \rightarrow 3 \rightarrow 2$                            & 0.29  \\ \hline
\end{tabular}
\caption{Progression sequences of first engagement of each type.  \label{tab:prog_seq}}
\end{table}

\section{Discussion and Conclusions}
\label{sec:conclusion}
In this paper, we have proposed a methodology for identifying tweets that criticise partisan news media, and we have analyzed how such tweets vary by user, news source, and time. Our classification experiments indicate that weak supervision can effectively train such a classifier with limited manually-annotated data.

Some of the substantive results are intuitive -- e.g., hyperpartisan users are more likely to criticise media from the other side (Figure~\ref{fig.dratio_user}), with CNN and Fox receiving the largest shares of critical tweets (Figure~\ref{fig.distrust_ns}). Other results are more nuanced -- e.g., unreliable news sources (-3, +3) do not necessarily receive more criticism than reliable news sources (-2, +2). Furthermore, we found substantial changes in critical tweets over time, including the tripling of the criticism ratio toward liberal media in mid-2017 and the doubling of the criticism ratio toward conservative media in late-2018. Finally, our accounting for media-oriented critical tweets reveals that user news engagement is not as politically diverse as one might otherwise expect.

\section{Limitations}
\label{sec:limitations}
The limitations of this work include the following:
\begin{packed_item}
\item User sample: As we aimed to identify users who engage with political news, the results should not be interpreted as representative of all of Twitter or the U.S. While most users appear to be based in the U.S., we made no effort to exclude users from other countries.

\item Media sample: While we considered a wide range of news sources, for our primary results we focus on engagement with partisan and unreliable sources, omitting media rated as -1, 0, or 1 by AllSides. As described in \S\ref{sec:problem}, this was done in part to focus our efforts on media most often mentioned in critical tweets. Future work should consider including additional news sources.

\item Article stance: While we attribute the overall stance of a news source to each of its articles, this assumption may not hold for all cases. Individual news articles can vary in their political leanings. Future work should consider analyzing the stances of news articles at a more granular level to account for this variability.

\item News source stance: This study assumes that the partisan stances of news sources are static over time. However, in reality, these stances may shift due to various factors such as changes in policies or broader political dynamics. Future work could assume dynamic partisan stances to better capture the evolving nature of news sources and their potential impact on user engagement.

\item Classifier noise: As our experiments indicate, the classifier is imperfect, and these errors can propagate to the analysis in \S\ref{sec:analysis}. Future work could apply adjustment methods to calibrate estimates of critical tweets~\cite{forman2005counting,keith2018uncertainty}.

\end{packed_item}

We acknowledge that we do not consider the issue of whether criticism of the media fosters greater levels of democracy but rather whether a key feature of democracy -- criticism in media -- might play a role in promoting the consumption of more polarizing news. That said, criticism of the media itself is accepted by the public as a key feature to improve journalism \cite{craft2016, cheruiyot2019}, a feature that is unlikely to be eliminated in online news consumption patterns anytime soon.

\section{Ethics Statement}
\label{sec:ethics}
 Understanding how hyperpartisanship and misinformation evolve over time has important implications for civic discourse and political engagement. This work analyzes public, historical data from Twitter users but does not intervene in any way. As such, the study was determined to be exempt by the institution's IRB committee. Nevertheless, inferences made from online data are error prone and, with improper use, could lead to disparate treatment or targeting of users. We have released tweet IDs for this dataset for research purposes, but not the raw content, which is in line with Twitter's terms of service and FAIR principles.

\section*{Acknowledgements}
This work was supported by the National Science Foundation Award \#1927407. AC was funded in part by the Tulane's Jurist Center for Artificial Intelligence and by Tulane's Center for Community-Engaged Artificial Intelligence.

\bibliography{references.bib}

\section*{Ethics Checklist}

\begin{enumerate}
\item For most authors...
\begin{enumerate}
    \item  Would answering this research question advance science without violating social contracts, such as violating privacy norms, perpetuating unfair profiling, exacerbating the socio-economic divide, or implying disrespect to societies or cultures?
    \answerYes{Yes, see Ethics section.}
  \item Do your main claims in the abstract and introduction accurately reflect the paper's contributions and scope?
    \answerYes{Yes}
   \item Do you clarify how the proposed methodological approach is appropriate for the claims made? 
    \answerYes{Yes, see Introduction.}
   \item Do you clarify what are possible artifacts in the data used, given population-specific distributions?
    \answerYes{Yes, see Limitations.}
  \item Did you describe the limitations of your work?
    \answerYes{Yes, see Limitations.}
  \item Did you discuss any potential negative societal impacts of your work?
    \answerYes{Yes, see Ethics.}
      \item Did you discuss any potential misuse of your work?
    \answerYes{Yes, see Ethics.}
    \item Did you describe steps taken to prevent or mitigate potential negative outcomes of the research, such as data and model documentation, data anonymization, responsible release, access control, and the reproducibility of findings?
    \answerYes{Yes, see Ethics.}
  \item Have you read the ethics review guidelines and ensured that your paper conforms to them?
    \answerYes{Yes}
\end{enumerate}

\item Additionally, if your study involves hypotheses testing...
\begin{enumerate}
  \item Did you clearly state the assumptions underlying all theoretical results?
    \answerYes{Yes, see Sections 6 and 7.}
  \item Have you provided justifications for all theoretical results?
    \answerYes{Yes, see Sections 6 and 7.}
  \item Did you discuss competing hypotheses or theories that might challenge or complement your theoretical results?
    \answerYes{Yes, see Limitations.}
  \item Have you considered alternative mechanisms or explanations that might account for the same outcomes observed in your study?
    \answerYes{Yes, see Limitations.}
  \item Did you address potential biases or limitations in your theoretical framework?
    \answerYes{Yes, see Limitations.}
  \item Have you related your theoretical results to the existing literature in social science?
    \answerYes{Yes, see Related Work.}
  \item Did you discuss the implications of your theoretical results for policy, practice, or further research in the social science domain?
    \answerYes{Yes, see Section 8.}
\end{enumerate}

\item Additionally, if you are including theoretical proofs...
\begin{enumerate}
  \item Did you state the full set of assumptions of all theoretical results?
    \answerNA{NA}
	\item Did you include complete proofs of all theoretical results?
    \answerNA{NA}
\end{enumerate}

\item Additionally, if you ran machine learning experiments...
\begin{enumerate}
  \item Did you include the code, data, and instructions needed to reproduce the main experimental results (either in the supplemental material or as a URL)?
    \answerNo{No, but have released code and data.}
  \item Did you specify all the training details (e.g., data splits, hyperparameters, how they were chosen)?
    \answerYes{Yes, see Appendix.}
     \item Did you report error bars (e.g., with respect to the random seed after running experiments multiple times)?
    \answerYes{Yes, see Section 6.}
	\item Did you include the total amount of compute and the type of resources used (e.g., type of GPUs, internal cluster, or cloud provider)?
    \answerYes{Yes, see Appendix.}
     \item Do you justify how the proposed evaluation is sufficient and appropriate to the claims made? 
    \answerYes{Yes, see Section 6.}
     \item Do you discuss what is ``the cost`` of misclassification and fault (in)tolerance?
    \answerYes{Yes, see Section 10.}
  
\end{enumerate}

\item Additionally, if you are using existing assets (e.g., code, data, models) or curating/releasing new assets...
\begin{enumerate}
  \item If your work uses existing assets, did you cite the creators?
    \answerYes{Yes, see Data.}
  \item Did you mention the license of the assets?
    \answerNA{NA}
  \item Did you include any new assets in the supplemental material or as a URL?
    \answerNo{Yes, we have released data and code at \url{https://github.com/tapilab/icwsm-2024-news-criticism}.}
  \item Did you discuss whether and how consent was obtained from people whose data you're using/curating?
    \answerYes{Yes, see Section 10.}
  \item Did you discuss whether the data you are using/curating contains personally identifiable information or offensive content?
    \answerYes{Yes, see Section 10.}
\item If you are curating or releasing new datasets, did you discuss how you intend to make your datasets FAIR?
\answerYes{Yes, see Section 10.}
\item If you are curating or releasing new datasets, did you create a Datasheet for the Dataset? 
\answerNo{See \url{https://github.com/tapilab/icwsm-2024-news-criticism}.}
\end{enumerate}

\item Additionally, if you used crowdsourcing or conducted research with human subjects...
\begin{enumerate}
  \item Did you include the full text of instructions given to participants and screenshots?
    \answerNA{NA}
  \item Did you describe any potential participant risks, with mentions of Institutional Review Board (IRB) approvals?
    \answerNA{NA}
  \item Did you include the estimated hourly wage paid to participants and the total amount spent on participant compensation?
    \answerNA{NA}
   \item Did you discuss how data is stored, shared, and deidentified?
    \answerNA{NA}
\end{enumerate}

\end{enumerate}

\appendix

\section{Bot Heuristics}
\label{sec:appendix_bot}

After our initial seed user collection, we filter these to remove suspected bot accounts as well as those likely to be celebrities or organizations. We use a set of heuristics from the literature for this filtering step~\cite{cresci2015fame}, where we look at different characteristics of each account and compare them against different cut-off values. The characteristics and their corresponding cut-off values are as follows 
\begin{enumerate}
    \item Follower Size ($\leq 1000$)
    \item Following Size ($\leq 1000$)
    \item Daily Tweet Activity ($\leq 10$)
    \item Total Tweets authored during the life of the account ($\geq 1000$ and $\leq 30000$)
\end{enumerate}

\subsection{Additional Validation}
\label{sec:validation}
To further validate the quality of the predictions the set of 1.2M tweets, we performed two additional checks. First, we manually annotated a random sample of 150 tweets predicted as criticism and 150 predicted as non-criticism, finding an accuracy score of 85\%, in line with the the results above. 

Next, we analyze the top terms in the tweets predicted to be critical and those predicted to be not critical. To do so, we computed chi-squared values for all terms to identify those that are significantly more likely in one class versus the other. For the critical class, the top terms are \{fake, propaganda, \#fakenews, lies, hey, tagged, news, FALSE, conspiracy, dear\}; for the non-critical class, the top terms are \{via, breaking, new, biden, says, bill, \#oann, senate, texas, federal\}. While many of these top terms are expectedly included in the heuristic functions (Table~\ref{tab:text_rules}), it is notable that many additional terms are discovered to predict critical tweets, including \{hey, tagged, dear, sponsors, misleading, coverup\}. Terms like ``hey'' and ``dear'' arise from tweets like ``Hey @cnn, fire your reporters.'' Combined with the accuracy results in Table~\ref{tab:model_perf}, these additional checks give us more confidence in the quality of the data annotated by the classifier.

\section{Labeling Functions}
\label{sec:appendix_lf}

In the following subsections we define and describe each of the labeling functions in detail.

\subsection{Based on User Profile ($\phi_{up}$)}

This labeling function first uses heuristics based on a user's historic news engagements and political accounts they follow to estimate their partisan lean. Then, it labels tweets that engage with news sources of the opposite partisan lean as critical, and those that engage with the same partisan lean as not critical. We denote this labeling function as $\phi_{up}$, where $ \{e^a_{up},e^{drt}_{up},e^{pf}_{up}\}$ is the collection of  heuristics it uses to assign a label to a given tweet, based on three different methods to estimate a user's partisan lean.

First, $e^a_{up}$ measures the partisan distribution of \textit{all}  historic tweets of user $u_j$ that mention a news source. It computes the fraction of tweets that engage with conservative $(c^{u_j}_{ac})$ and liberal $(c^{u_j}_{al})$ news sources. If the minimum of these fractions is less than or equal to a threshold $\delta_a$, then the heuristic labels all tweets that mentions a news source. 
It assigns a value of 1 (critical) if the tweet $t^{u_j}_i$ of the user engages with a news source whose stance $s^{u_j}_i$ is equal to the minority partisan stance $ps^a_{min}(u_j)$ of the user's news engagements and 0 (not critical) otherwise. For example, if the user is strongly conservative ($c^{u_j}_{ac} >> c^{u_j}_{al}$), then the minority stance would be liberal and vice versa. We also check to see if the total number of a user's news engagement tweets are greater than a threshold $\rho_a$ to make sure enough user information is present for the heuristic to work effectively.

\begin{table}[t]
\centering
\footnotesize
\begin{tabular}{cc}
\hline
\textbf{Threshold} & \textbf{Values}                                             \\ \hline
$\delta$                                                                                    & 0.05, 0.1, 0.15, 0.2, 0.25, 0.3, 0.35 \\
$\rho$                                                                                      & 5, 10, 15, 20, 25, 30              \\ \hline
\end{tabular}
\caption{Thresholds for $\phi_{up}$ tuned on validation data.}
  \label{tab:user_params}
\end{table}

\begin{table}[t]
\centering
\footnotesize
\begin{tabular}{|p{.04\textwidth}|p{0.32\textwidth}|p{.04\textwidth}|}
\hline
\multicolumn{1}{|c|}{\textbf{Name}} & \multicolumn{1}{c|}{\textbf{Heuristics}}  & \textbf{Label} \\ \hline
\multirow{2}{*}{$e^{d}_{tt}$}                 & cover up, covering up, shameful reporting, fakenews, fraud news, racist news, fraud network, racist network, not reporting,  untrusted news, shit news, half truths, tell the truth, cant handle the truth, bunch of crap, stop lying, brainwashed, misinformation, disinformation, exaggerations, scaremongering,  propaganda, fearmongering, hypocrisy, boycott                                                                                                                   & 1                    \\ \cline{2-3} 
                                              & watch this, must watch, live update, listen to, please read, read this, must read, worth reading, please share, study finds, top stories, top story, shocking news & 0                    \\ \hline
\multirow{2}{*}{$e^{c}_{tt}$}                 & 
(false, fake, hoax, fictitious, misrepresent, one sided, bullshit, crap, shit, garbage, exaggerate) $\wedge$ (news, stories, reporting, narrative, media, network, reports) \newline
(conspiracy) $\wedge$ (theories, theory) \newline 
(misinform, mislead) $\wedge$ (public,people,america) \newline 
(made up, make up) $\wedge$ (lies, crap, shit)\newline 
(brainwash, deceive) $\wedge$ (people, public, america)\newline 
(spread) $\wedge$ (lies, propaganda, conspiracies, shit, fear)\newline 
(biased) $\wedge$ (news, report, narrative, network, media, shit)
& 1                    \\ \cline{2-3} 
                                              & 
(breaking) $\wedge$ (news, exclusive, report, story) \newline 
(watch) $\wedge$ (now, live) \newline
(good, great, inspiring, incredible, real, thanks, thx, latest, fantastic) $\wedge$ (news, report, story, journalism, narrative, article, piece, video) \newline
(best) $\wedge$ (news, report, video)                             & 0                    \\ \hline
\multirow{2}{*}{$e^{p}_{tt}$}                 & expose @NS, exposing @NS, exposes @NS, @NS exposed, @NS sucks, @NS is a joke, @NS fuck you, fuck you @NS, screw you @NS, @NS screw you, fuck @NS, @NS crap, @NS is crap, crap from @NS, @NS should fire, cant trust @NS, can not trust @NS, dont trust @NS,  do not trust @NS     
& 1                    \\ \cline{2-3} 
                                              & via @NS                   & 0                    \\ \hline
\end{tabular}
\caption{Text heuristics ($\phi_{tt}$)}
  \label{tab:text_rules}
\end{table}

\begin{equation}
    e^a_{up}(t^u_i) = 
\begin{cases}
    -1 ~~\text{if } \hbox{number of news tweets} < \rho_a \\ \hbox{~~~~~~~~or } \textbf{min}(c^{u_j}_{ac},c^{u_j}_{al}) > \delta_a\\

    ~~~1 ~~\text{if } s^{u_j}_i = {ps}^a_{min}(u_j)  \\
    
    ~~~0 ~~\text{otherwise}
\end{cases} 
\end{equation}

Next, $e^{drt}_{up}$ represents a similar heuristic as $e^a_{up}$ but instead of considering all the tweets where the user engages with a news source, we only consider \textit{direct retweets} (these are retweets whose original author is an official Twitter account of a news source). Focusing on direct retweets reduces noise introduced by quote tweets or news source mentions, which may be critical. Analogous thresholds $\delta_{drt}$ and $\rho_{drt}$ are used to ensure that the minority partisan class is small enough, as well as to ensure that the overall volume of direct retweets is sufficient to estimate a user's partisan lean.

The third heuristic $e^{pf}_{up}$ is analogous to the above, but instead estimates a user's partisan lean based on the set of political Twitter accounts a user follows. Using a dataset of politician Twitter accounts with party affiliation\footnote{\url{https://www.propublica.org/datastore/dataset/politicians-tracked-by-politwoops}}, it follows the same procedure as above to compute the fraction of liberal and conservative accounts a user follows, again assigning tweets as critical if they are cross-partisan according to the estimated user stance, and 0 otherwise. Thresholds $\delta_{pf}$ and $\rho_{pf}$ are used for this heuristic.

The final labeling function $\phi_{up}$ uses a unanimous vote to assign the final labels for a user's tweet $t^u_i$. 

\begin{equation}
    \phi_{up}(t^{u_j}_i) = 
\begin{cases}
    1,~\text{if }  e^a_{up}(t^{u_j}_i) == e^{drt}_{up}(t^{u_j}_i) \\ ~~~== e^{pf}_{up}(t^{u_j}_i) == 1 \\ 
    0,~\text{if } e^a_{up}(t^{u_j}_i) == e^{drt}_{up}(t^{u_j}_i) \\ ~~~== e^{pf}_{up}(t^{u_j}_i) == 0 \\
    -1,~otherwise
\end{cases}  
\end{equation}

The threshold parameters ($\delta, \rho$) are tuned across different ranges (Table \ref{tab:user_params}) to maximize accuracy on the manually annotated validation set. The final settings are: $\delta_{a}=.1$, $\delta_{drt}=.15$, $\delta_{pf}=.05$ and $\rho_a=10$, $\rho_{drt}=25$, $\rho_{pf}=5$.

\subsection{Based on Tweet Text ($\phi_{tt}$)}
This labeling function uses multiple text based heuristics to  annotate tweets (Table~\ref{tab:text_rules}). We denote this labeling function as $\phi_{tt}$, where $\{e^{d}_{tt}, e^{c}_{tt}, e^{p}_{tt}\}$ represents the types of heuristics it uses to assign a label to a given tweet. 

Here $e^{d}_{tt}$ represents a set of heuristics that perform a direct string match to label both the critical ($e^{d+}_{tt}$) and non-critical classes ($e^{d-}_{tt}$). It labels a user's tweet $t^{u_j}_i$ if any of these keywords or phrases are present in the text of the tweet.

$e^{c}_{tt}$ represents a set of word collocation heuristics, each of which is a conjunction of two disjunctive clauses: $\{w_1^a \vee \ldots \vee w_k^a\} \wedge \{w^b_1 \vee \ldots \vee w^b_m\}$, e.g., $\{$fake $ \vee$ false$\} \wedge \{$news $\vee$ media $\vee$ stories$\}$.
Similar to $e^{d}_{tt}$, this heuristic set contains different sets of rules for both the critical ($e^{c+}_{tt}$) and non-critical ($e^{c-}_{tt}$) class.

Lastly, $e^{p}_{tt}$ represents a set of phrase heuristics which check to see if a news source is mentioned with certain neighboring words in a specific order.

The final labeling function $\phi_{tt}$ uses a "logical or" to assign the final labels for a user's tweet $t^{u_j}_i$.

\begin{equation}
    \phi_{tt}(t^{u_j}_i) = 
\begin{cases}
    1, ~\text{if }  e^{d+}_{tt}(t^{u_j}_i) == 1 \\ ~~~~\lor e^{c+}_{tt}(t^{u_j}_i) == 1 \\ ~~~~\lor e^{p+}_{tt}(t^{u_j}_i) == 1 \\ 
    0, ~\text{if }  e^{d-}_{tt}(t^{u_j}_i) == 0 \\ ~~~~\lor e^{c-}_{tt}(t^{u_j}_i) == 0 \\ ~~~~\lor e^{p-}_{tt}(t^{u_j}_i) == 0 \\
    -1,~otherwise
\end{cases}  
\end{equation}

If any of these heuristics fail to assign a label due to them not being present in the text of tweet or if the heuristics for both the critical and non-critical classes both fire, we assign a label of -1. For the complete list of heuristics see Table \ref{tab:text_rules}.

\subsection{Union of Labeling Functions ($\phi_{un}$)}

This is a labeling function that performs a union over the labels generated by $\phi_{up}$ and $\phi_{tt}$. It assigns a label of 1 if either of the labeling functions assigns a label of 1 and they don't conflict (similarly for class 0). It assigns a label of -1 when both labeling functions assign a label of -1 or they have conflicting labels (i.e $\phi_{up}(t^{u_j}_i) \ \ne \  \phi_{tt}(t^{u_j}_i)$). We denote this labeling function as $\phi_{un}$.

\section{User Features}
\label{sec:user_features}
The features for the User Network are described in Table \ref{tab:user_feats}.

\begin{table}
\centering
\footnotesize
\begin{tabular}{|p{0.16\textwidth}p{0.24\textwidth}|}
\hline
\textbf{Name}                                        & \textbf{Description}  \\ \hline
News Source Engagement Partisan Distribution         & The distribution of partisan stances of all the news sources the user engages with across all his tweets \\ \hline
Followed Politician Accounts Partisan Distribution   & The distribution of partisan stances of all politicians the user follows \\ \hline
Followed News Source Accounts Partisan Distribution  & The distribution of partisan stances of all news sources the user follows                                   \\ \hline
Engaged News Source Partisan Stance                  &  The partisan stance of the news source the  user currently engages with in the current tweet            \\ \hline
Tweet Type                                           & The type of tweet (i.e retweet, replied\_to, status, quote)                                                 \\ \hline
News Source Engagement Type                          & How the user engages with the news source in the current tweet (i.e mention, URL)                           \\ \hline
Is Direct Reply                                      & If the current tweet is a direct reply to a news source twitter account                                  \\ \hline
Multiple News Source Engagement                      & If multiple news sources are mentioned in the current tweet                                                 \\ \hline
Public Metrics                                       & The public metrics of the current tweet \\ \hline
Engaged News Source Fraction                         & The fraction of engagements of the current news source engaged in the tweet                                 \\ \hline
\end{tabular}
\caption{User Based Features}
  \label{tab:user_feats}
\end{table}

\begin{table}
\centering
\footnotesize
  \begin{tabular}{ll}
  \hline
\textbf{Hyperparameter}    & \textbf{Values}         \\
\hline
Learning Rate              & 1e-2 to 1e-6            \\
Epochs                     & 50
\\
Early Stopping Patience    & 3,5,7                   \\
Hidden Units               & 64,128,256,512,1024 \\
Pre-trained Freezing       & True, False             \\
Hidden Activations         & Relu, Sigmoid           \\
Batch Size                 & 8,16,64,128,256         \\
Dropout                    & 0.05,0.1,0.2,0.3     \\
\hline
\end{tabular}
  \caption{Hyperparameter Values for Experiments}
  \label{tab:hyper}
\end{table}

\section{Experimental Settings}
Table~\ref{tab:hyper} shows all hyperparameters, optimized on the held-out validation set. The best setting is then used to compute accuracy on the test set (Table~\ref{tab:model_perf}). All experiments use a system with 4 Nvidia A5000 GPUs, 512 GB RAM and an AMD Ryzen Threadripper 3975WX CPU. We report mean and standard deviation of scores across five random seeds.

\section{Additional Performance Metrics of the Classification Models}

Table \ref{tab:add_model_perf} shows other measures of classification performance on the test set for all considered models.

\begin{table*}[h]
\small
\centering
\begin{tabular}{lcccc}
\hline
\multicolumn{1}{c}{\textbf{Network}} & \textbf{Accuracy}                              & \textbf{F1}                                    & \textbf{Precision}                             & \textbf{Recall}                                                               \\ \hline
Text Network + $\phi_{up}$           & 0.800 $\pm$ 0.049                              & 0.800 $\pm$ 0.028                              & 0.805 $\pm$ 0.023                              & 0.801 $\pm$ 0.026                                                            \\
Text Network + $\phi_{tt}$           & 0.723 $\pm$ 0.008                              & 0.717 $\pm$ 0.007                              & 0.738 $\pm$ 0.016                              & 0.723 $\pm$ 0.009                                                            \\
Text Network + $\phi_{un}$           & 0.813 $\pm$ 0.015                              & 0.810 $\pm$ 0.017                              & 0.825 $\pm$ 0.010                              & 0.813 $\pm$ 0.015                                                            \\

Text Network + DS           & \multicolumn{1}{l}{\textbf{0.840 $\pm$ 0.007}} & \multicolumn{1}{l}{\textbf{0.840 $\pm$ 0.007}} & \multicolumn{1}{l}{\textbf{0.840 $\pm$ 0.007}} & \multicolumn{1}{l}{\textbf{0.840 $\pm$ 0.007}}  \\

Text Network + IBCC           & \multicolumn{1}{l}{0.824 $\pm$ 0.015} & \multicolumn{1}{l}{0.824 $\pm$ 0.015} & \multicolumn{1}{l}{0.828 $\pm$ 0.017} & \multicolumn{1}{l}{0.824 $\pm$ 0.015}  \\

Text Network + EBCC           & \multicolumn{1}{l}{0.837 $\pm$ 0.006} & \multicolumn{1}{l}{0.837 $\pm$ 0.006} & \multicolumn{1}{l}{0.839 $\pm$ 0.006} & \multicolumn{1}{l}{0.837 $\pm$ 0.006}  \\

Text Network + DP      & \multicolumn{1}{l}{0.812 $\pm$ 0.013}          & \multicolumn{1}{l}{0.812 $\pm$ 0.013}          & \multicolumn{1}{l}{0.812 $\pm$ 0.012}          & \multicolumn{1}{l}{0.812 $\pm$ 0.013}                   \\
User Network + $\phi_{up}$           & 0.747 $\pm$ 0.049                              & 0.747 $\pm$ 0.050                              & 0.751 $\pm$ 0.047                              & 0.747 $\pm$ 0.050                                                           \\
User Network + $\phi_{tt}$           & 0.662 $\pm$ 0.018                              & 0.643 $\pm$ 0.033                              & 0.699 $\pm$ 0.022                              & 0.662 $\pm$ 0.018                                                           \\
User Network + $\phi_{un}$           & 0.746 $\pm$ 0.025                              & 0.745 $\pm$ 0.025                              & 0.747 $\pm$ 0.026                              & 0.746 $\pm$ 0.025                                                            \\
Combined Network + $\phi_{up}$       & 0.812 $\pm$ 0.014                              & 0.811 $\pm$ 0.015                              & 0.816 $\pm$ 0.013                              & 0.812 $\pm$ 0.014                                                            \\
Combined Network + $\phi_{tt}$       & 0.728 $\pm$ 0.006                              & 0.719 $\pm$ 0.009                              & 0.755 $\pm$ 0.019                              & 0.728 $\pm$ 0.006                                                            \\
Combined Network + $\phi_{un}$       & 0.799 $\pm$ 0.007                              & 0.797 $\pm$ 0.007                              & 0.808 $\pm$ 0.008                              & 0.799 $\pm$ 0.007                                                            \\
Combined Network + DS       & \multicolumn{1}{l}{0.816 $\pm$ 0.014}          & \multicolumn{1}{l}{0.818 $\pm$ 0.015}          & \multicolumn{1}{l}{0.816 $\pm$ 0.014}          & \multicolumn{1}{l}{0.816 $\pm$ 0.014}                   \\

Combined Network + IBCC  & \multicolumn{1}{l}{0.785 $\pm$ 0.035}          & \multicolumn{1}{l}{0.783 $\pm$ 0.037}          & \multicolumn{1}{l}{0.792 $\pm$ 0.032}          & \multicolumn{1}{l}{0.785 $\pm$ 0.036}                    \\ 

Combined Network + EBCC  & \multicolumn{1}{l}{0.810 $\pm$ 0.023}          & \multicolumn{1}{l}{0.810 $\pm$ 0.023}          & \multicolumn{1}{l}{0.810 $\pm$ 0.023}          & \multicolumn{1}{l}{0.810 $\pm$ 0.022}                    \\ 

Combined Network + DP  & \multicolumn{1}{l}{0.826 $\pm$ 0.014}          & \multicolumn{1}{l}{0.826 $\pm$ 0.014}          & \multicolumn{1}{l}{0.826 $\pm$ 0.015}          & \multicolumn{1}{l}{0.826 $\pm$ 0.014}                   \\ \hline
\end{tabular}
  \caption{Test set Performance for combinations of model, labeling function, and label denoising methods.}
  \label{tab:add_model_perf}
\end{table*}

\section{Change in Criticism of Popular News Sources }
\label{sec:appendix_dis_ns}
\begin{figure}[ht]
    \centering
    \includegraphics[width=\linewidth]{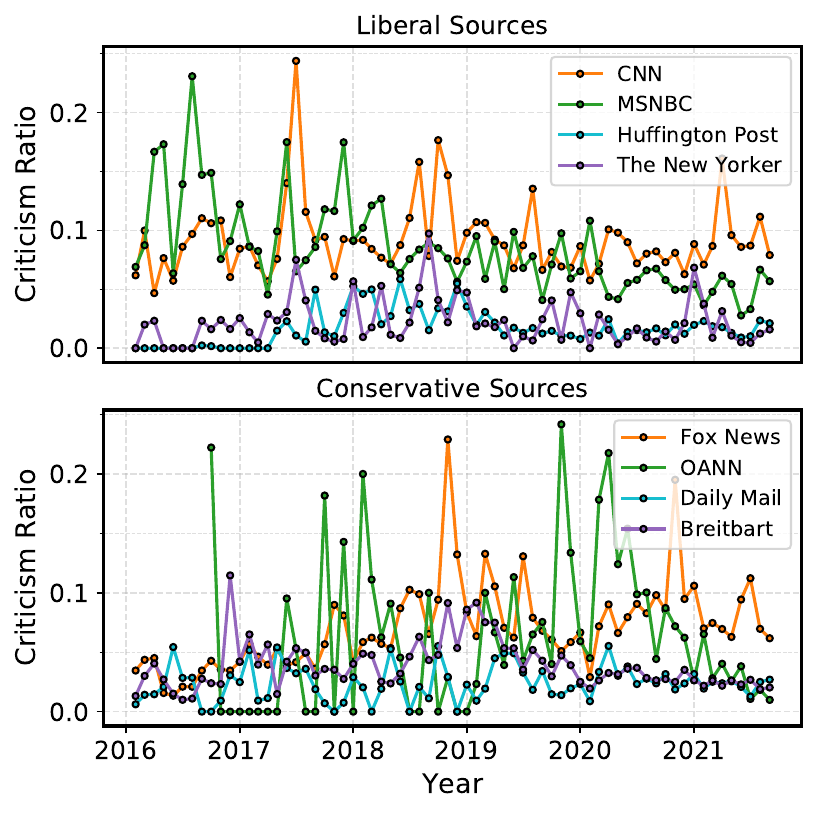}
    \caption{Criticism of Popular News Sources through Time}
    \label{fig.distrust_time_ns}
\end{figure}

Based on the results from Figure \ref{fig.distrust_ns}, we also analyzed how the criticism shown towards these news sources changed through time. The resulting graph can be seen in Figure \ref{fig.distrust_time_ns}.

\end{document}